\documentclass[pdflatex,sn-basic,iicol,Numbered]{sn-jnl_custom}

\usepackage{graphicx}%
\usepackage{multirow}%
\usepackage{amsmath,amssymb,amsfonts}%
\usepackage{dblfloatfix}
\usepackage{amsthm}%
\usepackage{mathrsfs}%
\usepackage[title]{appendix}%
\usepackage{adjustbox}
\usepackage{cuted} % nel preambolo

\usepackage{xcolor}%
\usepackage{textcomp}%
\usepackage{manyfoot}%
\usepackage{booktabs}%
\usepackage{url}
\usepackage{algorithm}%
\usepackage{algorithmicx}%
\usepackage{algpseudocode}%
\usepackage{listings}%
\usepackage{verbatim}
\usepackage{tikz}
\usetikzlibrary{quantikz2}
\usepackage{subcaption} % For subfigures
\usepackage{placeins}
\usepackage{float}
\usepackage{amsfonts} 
\usepackage[normalem]{ulem}

\theoremstyle{thmstyleone}%
\theoremstyle{thmstyletwo}%

\theoremstyle{thmstylethree}%
\usepackage{graphicx}
\begin{document}

%%=============================================================%%
%% GivenName	-> \fnm{Joergen W.}
%% Particle	-> \spfx{van der} -> surname prefix
%% FamilyName	-> \sur{Ploeg}
%% Suffix	-> \sfx{IV}
%% \author*[1,2]{\fnm{Joergen W.} \spfx{van der} \sur{Ploeg} 
%%  \sfx{IV}}\email{iauthor@gmail.com}
%%=============================================================%%

%%
%% The "title" command
\title{Quantum Approaches to Urban Logistics: From Core QAOA to Clustered Scalability}

%%
%% The "author" command and its associated commands are used to define
%% the authors and their affiliations.
\author*[1]{\fnm{Fabio} \sur{Picariello}}\email{fabio.picariello@eng.it}

\affil[1]{Eng AI \& Data @ Engineering Group, Piazzale dell'Agricoltura 24, 00144 Roma, Italy}

\author*[2]{\fnm{Gloria} \sur{Turati}}\email{gloria.turati@polimi.it}

\affil[2]{Politecnico di Milano, Piazza Leonardo da Vinci 32, 20133 Milano, Italy}

\author[1]{\fnm{Riccardo} \sur{Antonelli}}
\author[1]{\fnm{Igor} \sur{Bailo}}
\author[1]{\fnm{Susanna} \sur{Bonura}}
\author[1]{\fnm{Gianmarco} \sur{Ciarfaglia}}
\author[1]{\fnm{Salvatore} \sur{Cipolla}}
\author[2]{\fnm{Paolo} \sur{Cremonesi}}
\author[2]{\fnm{Maurizio} \sur{Ferrari Dacrema}}
\author[1]{\fnm{Michele} \sur{Gabusi}}
\author[3]{\fnm{Ivan} \sur{Gentile}}
\author[1]{\fnm{Vito} \sur{Morreale}}
\author[1]{\fnm{Antonio} \sur{Noto}}

\affil[3]{International Foundation Big Data And Artificial Intelligence For Human Development
Via Galliera n. 32 – Bologna}

%%==================================%%
%% Sample for unstructured abstract %%
%%==================================%%

\abstract{The Traveling Salesman Problem (TSP) is a fundamental challenge in combinatorial optimization, widely applied in logistics and transportation. As the size of TSP instances grows, traditional algorithms often struggle to produce high-quality solutions within reasonable timeframes. This study investigates the potential of the Quantum Approximate Optimization Algorithm (QAOA), a hybrid quantum-classical method, to solve TSP under realistic constraints. We adopt a QUBO-based formulation of TSP that integrates real-world logistical constraints reflecting operational conditions—such as vehicle capacity, road accessibility, and time windows—while ensuring compatibility with the limitations of current quantum hardware. Our experiments are conducted in a simulated environment using high-performance computing (HPC) resources to assess QAOA's performance across different problem sizes and quantum circuit depths. In order to improve scalability, we propose \textit{clustering QAOA} (Cl-QAOA), a hybrid approach combining classical machine learning with QAOA. This method decomposes large TSP instances into smaller sub-problems, making quantum optimization feasible even on devices with a limited number of qubits.
The results offer a comprehensive evaluation of QAOA's strengths and limitations in solving constrained TSP scenarios. This study advances quantum optimization and lays groundwork for future large-scale applications.}


%%\pacs[JEL Classification]{D8, H51}

%%\pacs[MSC Classification]{35A01, 65L10, 65L12, 65L20, 65L70}

\maketitle

\section{Introduction}

Optimization in logistics is a critical component across a wide range of industrial domains, including supply chain management and urban delivery systems. Many of the underlying problems, such as the well-known Traveling Salesman Problem (TSP)~\citep{applegate2006traveling}, belong to the class of NP-hard problems, for which exact algorithms become computationally infeasible as the problem size increases~\citep{woeginger2003exactMethods}.

To address the complexity of these problems, a variety of heuristic and metaheuristic approaches have been developed~\citep{boussaid2013metaheuristic}, capable of producing high-quality approximate solutions within reasonable time frames. However, as problem size increases, these algorithms often yield solutions that diverge from the global optimum.

Recent advances in quantum computing have opened new avenues for tackling large-scale combinatorial optimization problems. Quantum algorithms leverage phenomena like superposition and entanglement to explore vast solution spaces, potentially outperforming classical methods in efficiency. By formulating these problems in a quantum framework, it becomes possible to seek high-quality approximate solutions while potentially reducing computational overhead.

This work focuses on the Quantum Approximate Optimization Algorithm (QAOA)~\citep{farhi2014qaoa, blekos2023qaoa}, a variational quantum algorithm particularly suited for near-term quantum hardware. Tailored for discrete optimization tasks~\citep{crooks2018qaoa, willsch2020qaoa, cook2020qaoa, brandhofer2022qaoa, tabi2020qaoa, lin2016qaoa, kurowski2023qaoa, turati2022qaoa}, QAOA operates through an iterative interplay between quantum evolution and classical parameter optimization.

In this study, we investigate the application of QAOA to TSP instances enriched with realistic logistical constraints inspired by real-world transportation network limitations. We assess the algorithm's performance on both synthetic benchmarks and real-world datasets, evaluating solution quality and scalability.

Moreover, given the current limitations of quantum hardware in terms of available logical qubits, we investigate the integration of QAOA with machine learning (ML) techniques to reduce the quantum resource overhead. This hybrid approach aims to enable the solution of larger problem instances even with today's constrained quantum technologies.

The main contributions of this work are as follows:
\begin{itemize}
    \item The implementation of a Grover-inspired mixer to enforce the canonical one-city-per-step constraint within the QAOA framework.
    \item The formulation of domain-specific constraints in a manner compatible with quantum hardware.
    \item An empirical evaluation of QAOA's solution quality on realistic, data-driven logistics scenarios.
    \item The design of a hybrid architecture that combines clustering-based machine learning techniques with QAOA, enabling scalability to large TSP instances.
    \item A comprehensive temporal scaling analysis of both QAOA for the TSP and our hybrid machine learning–augmented method, assessing their practical applicability and scalability.
\end{itemize}

The remainder of the paper is organized as follows. \textbf{Section~\ref{sec:background}} introduces the TSP and the QAOA, together with the encoding strategies adopted for quantum optimization. \textbf{Section~\ref{sec:methodology}} presents the proposed methodology, including the modeling of real-world constraints compatible with current quantum hardware limitations, the choice of the QAOA mixer, the detailed structure of our clustering-based QAOA approach, and the temporal scaling behavior of both QAOA and its clustering-enhanced version. \textbf{Section~\ref{sec:experimental_setup}} describes the experimental setup, including the datasets, the computational environment—comprising both a local high-performance workstation and the CINECA supercomputing infrastructure\footnote{\url{https://leonardo-supercomputer.cineca.eu/}}—the evaluation metrics, and the algorithmic configurations employed. \textbf{Section~\ref{sec:results}} reports the experimental findings, focusing on solution quality and computational time across different algorithms and configurations, while \textbf{Section~\ref{sec:discussion}} interprets these results. Finally, \textbf{Section~\ref{sec:conclusions}} summarizes the main contributions and outlines future research directions.

This paper extends our previous work presented at the International Workshop on AI for Quantum and Quantum for AI (AIQxQIA 2025), part of ECAI 2025, titled ``QAOA for Efficient Urban Logistical Ecosystem''.

\section{Background and Related Work}
\label{sec:background}

This section outlines the theoretical and methodological foundations of this study. We begin by reviewing the classical TSP and the QAOA. We then introduce a Quadratic Unconstrained Binary Optimization (QUBO) formulation of the TSP, which enables its integration into the QAOA framework for quantum optimization.

\subsection{The Traveling Salesman Problem (TSP)}

The TSP~\citep{applegate2006traveling} is a foundational problem in combinatorial optimization, with extensive real-world applications in areas such as vehicle routing, logistics, circuit design, and scheduling ~\citep{cattelan2022vrpqubo, azad2023vrp, mohanty2024vrp, radzihovsky_2019}. In its classical form, the problem requires finding the shortest route that visits each city once and returns to the starting point. We refer to these conditions as \textit{canonical constraints}.

Although the classical TSP formulation captures the essence of many routing problems, it may not fully address the practical constraints encountered in real-world scenarios. Practical applications often impose additional \textit{logistical constraints}, including:
\begin{itemize}
    \item \textbf{Capacity constraints}, which limit the load or service capacity of vehicles.
    \item \textbf{Time windows}, requiring visits to occur within specific time intervals.
    \item \textbf{Road accessibility}, which restricts the use of certain paths or routes.
\end{itemize}

Incorporating these real-world constraints leads to a significantly more complex version of the TSP, rendering it even more computationally challenging to solve. 

Advances in quantum computing have opened new directions for addressing such complex optimization problems~\citep{irie2019quantum}. Both quantum annealing and variational quantum algorithms have been investigated for solving symmetric, asymmetric, and constrained TSP variants. For example, quantum annealing has been employed to tackle TSP instances by encoding them into a QUBO form and mapping them, also onto quantum hardware such as D-Wave systems~\citep{silva2021mapping}. These approaches have shown promising results for small- to medium-sized problem instances, particularly when constraints are incorporated directly into the cost function~\citep{warren2020solving}.

Although quantum annealing dominated early quantum optimization research, gate-based quantum algorithms—compatible with universal quantum computers—are gaining increasing attention in this field~\citep{lytrosyngounis2025hybrid}.
While fully quantum solutions still trail behind state-of-the-art classical solvers, hybrid approaches significantly narrow the performance gap and offer scalable alternatives for future applications.

Gate-based quantum algorithms are particularly suitable for constrained variants of TSP, as they allow flexible encoding of problem-specific constraints directly into the quantum circuit.

\subsection{Quantum Approximate Optimization Algorithm (QAOA)}

The QAOA~\citep{farhi2014qaoa} is a hybrid quantum-classical algorithm designed to address combinatorial optimization problems. Its shallow circuit depth and robustness to noise make it particularly suitable for near-term quantum devices~\citep{yanakiev2023dynamic}.

QAOA constructs a parameterized quantum state by alternating two types of unitary operators: one derived from the cost Hamiltonian $H_c$, which encodes the objective function, and another from the mixing Hamiltonian $H_m$, which enables exploration of the solution space. These operators are applied in $p$ alternating layers, where $p$ governs the trade-off between approximation quality and circuit complexity.

Starting from an initial quantum state $|s\rangle$, the system evolves under the influence of parameter vectors $\vec{\gamma} = (\gamma_1, \dots, \gamma_p)$ and $\vec{\beta} = (\beta_1, \dots, \beta_p)$, resulting in the final quantum state:
\begin{equation}
\label{eq:qaoa_evolution}
    |\vec{\gamma}, \vec{\beta} \rangle =  \prod_{j=1}^{p} e^{-i\beta_jH_m}e^{-i\gamma_j H_c}|s\rangle.
\end{equation}
This state is measured repeatedly to estimate the the expectation value of the cost Hamiltonian:
\begin{equation}
\label{expectation_QAOA}
    \langle H_C\rangle_{\gamma,\beta} = \langle \vec{\gamma}, \vec{\beta} | H_c | \vec{\gamma}, \vec{\beta} \rangle,
\end{equation}
which is then minimized by a classical optimizer that iteratively adjusts $\vec{\gamma}$ and $\vec{\beta}$. The goal is to steer the quantum state toward the ground state of $H_c$, corresponding to the optimal solution.

For unconstrained problems, a common choice for the mixing Hamiltonian is the $X$-Mixer:
\begin{equation}
    H_m = \sum_{i=1}^{n} X_i,
\end{equation}
where $X_i$ denotes the Pauli-X operation on the qubit $i$.
This mixer enables exploration across the entire solution space.

In contrast, constrained problems benefit from alternative mixers such as the Grover mixer~\citep{bartschi2020grover}. This approach initializes the quantum system in a uniform superposition over feasible solutions and employs a mixer that restricts transitions to the feasible subspace. Let $U_s$ be a unitary operator that maps the all-zero state to a uniform superposition over the feasible set $F$:
\begin{equation}
    U_s |0\rangle^{\otimes n} = \frac{1}{\sqrt{|F|}} \sum_{x \in F} |x\rangle.
\end{equation}
The Grover mixer is then defined as:
\begin{equation}
    U_m(\beta_i) = U_s \left( I_d - \left(1 - e^{-i \beta_i}\right) |0\rangle \langle 0| \right) U_s^\dagger. \label{mixer_operator}
\end{equation}
This operator replaces $e^{-i\beta_i H_m}$ in Equation~\ref{eq:qaoa_evolution} for each $\beta_i$, ensuring that the quantum evolution remains confined to the feasible subspace. This restriction reduces the effective search space and enhances trainability by simplifying the optimization landscape.

QAOA offers several advantages: it is versatile, compatible with NISQ-era hardware~\citep{preskill2018nisq}, and its performance can be improved by increasing the circuit depth $p$. However, its effectiveness depends on the initialization of parameters and the complexity of the classical optimization, particularly when strong constraints are embedded in $H_c$.

A notable class of problems solvable via QAOA is the family of QUBO problems~\citep{glover2022qubo}, which are NP-hard and take the form:
\begin{equation}
    \min_{x \in \{0,1\}^n} x^\top Q x,
\end{equation}
where $x \in \{0,1\}^n$ is a binary vector and $Q \in \mathbb{R}^{n \times n}$ is a symmetric (or upper triangular) matrix defining the cost landscape.

QUBO problems can be reformulated as ground state search problems using the Ising model representation~\citep{lucas2014ising}, making them amenable to quantum algorithms such as QAOA. In this representation, the Hamiltonian is diagonal in the computational basis, so its ground state corresponds to a basis state. Thus, the measurement yields a bitstring encoding a candidate solution to the original QUBO problem.

\subsection{QAOA Encoding for the TSP}

The QAOA can be adapted to solve TSP instances by reformulating it as a QUBO model~\citep{cattelan2022vrpqubo}. In this formulation, binary variables indicate whether a specific city is visited at a particular time step in the tour. The objective function, typically defined in terms of travel distance or time, is minimized within the QAOA framework.

The TSP is governed by canonical constraints: each time step must correspond to exactly one visited city, and each city must be visited exactly once. Since QUBO models are inherently unconstrained, these conditions are enforced by incorporating penalty terms into the cost function, which penalize infeasible solutions.

Consider a TSP instance with $n$ cities. The QUBO formulation introduces $n^2$ binary variables, where each variable $x_{i,t}$ indicates whether city $i$ is visited at time step $t$:
\begin{equation}
x_{i,t} =
\begin{cases}
1 & \text{if city } i \text{ is visited at step } t, \\
0 & \text{otherwise}.
\end{cases}
\end{equation}

The cost Hamiltonian encoding the objective function is defined as:
\begin{equation}
    D(x) = \sum_{i,j=1}^{n} \omega_{i,j} \sum_{t=1}^{n-1} x_{i,t} x_{j,t+1}, \label{cost_H_QUBO}
\end{equation}
where $\omega$ is the cost matrix, with entries $\omega_{i,j}$ representing the cost of traveling from city $i$ to city $j$.

To enforce the TSP constraints, we introduce penalty terms that discourage invalid configurations:
\begin{align}
\label{eq:penalty_1}
    P(x) =&  \lambda_{p} 
        \underbrace{\sum_i \left( \sum_t x_{i,t} - 1 \right)^2}_{\text{\small each city visited once}}\\ & \nonumber   
    + \lambda_{q}  
        \underbrace{\sum_t \left( \sum_i x_{i,t} - 1 \right)^2}_{\text{\small one city per time step}},
\end{align}
where $\lambda_{p}$ and $\lambda_{q}$ are penalty weights that increase the cost of infeasible solutions while leaving the cost of feasible solutions unchanged. Their values are typically selected so that constraint violations are sufficiently penalized compared to the variability of the objective term, and their magnitude depends on the specific instance and modeling choices.

In our approach, the second constraint (one city per time step) is not enforced through penalty terms but instead via the Grover mixer mechanism, as described in Subsection~\ref{subsec:grover_mixer}. Therefore, the only penalty term explicitly included in the cost function is:
\begin{equation}
\label{eq:penalty_2}
    \tilde{P}(x) = \lambda_{p} 
        \sum_i \left( \sum_t x_{i,t} - 1 \right)^2,
\end{equation}
which ensures that each city is visited exactly once.

The final cost function $C(x)$ for a TSP instance without additional logistical constraints is thus given by:
\begin{equation}
\label{eq:cost_2}
    C(x) = D(x) + \tilde{P}(x).
\end{equation}

\section{Methods}
\label{sec:methodology}

This section describes the methodology of our study. In particular, we discuss the modeling of logistical constraints compatible with quantum hardware, the Grover-inspired mixer, the detailed structure of our clustering-based QAOA approach, and the temporal scaling behavior of both standard QAOA and Cl-QAOA for large problem instances.

\subsection{Logistical Constraints}
\label{subsec:logistical_constraints}

In order to enhance the relevance and applicability of quantum algorithms to logistics and routing problems, it is crucial to incorporate such constraints into the quantum formulation. This must be done carefully to avoid excessive increases in circuit depth, qubit count, and overall algorithmic complexity.

The constraints adopted in this work are inspired by prior formulations~\citep{irie2019quantum, papalitsas2019constraints}, but have been adapted into a binary representation to facilitate quantum implementation and reduce resource overhead.

\subsubsection{Binary Node Compatibility}

To simulate a capacitated TSP within the limitations of current quantum hardware and simulators, we propose a simplified formulation that approximates capacity constraints in a tractable manner, labeled Binary Node Compatibility (BNC).

We define a binary vector $k \in \{0,1\}^n$, where each component $k_i$ encodes the categorical state of node (or city) $i$:
\begin{equation}
    k_i = 
    \begin{cases}
        1 & \text{if node } i \text{ belongs to category A}, \\
        0 & \text{if node } i \text{ belongs to category B}.
    \end{cases}
\end{equation}
This abstraction captures scenarios where nodes are partitioned into two distinct classes, and transitions between nodes of the same class incur a penalty. Representative examples may include electric vehicle routing—where customer visits (1) alternate with charging stops (0) to maintain operational feasibility—and waste collection, where pickup points (1) alternate with depot unloading locations (0).

The penalty term can be modeled as:

\begin{equation}
    K(x) = \lambda_k \sum_{\substack{i,j,t \\ i \neq j}} x_{i,t} x_{j,t+1} (1 - k_i \oplus k_j),
    \label{eq:capacity_constraints}
\end{equation}
where $\oplus$ denotes the binary XOR operation.

Notably, Equation~\ref{cost_H_QUBO} shares the same structural form as Equation \ref{eq:capacity_constraints}. Therefore, the BNC constraint can be directly embedded into the cost matrix, without increasing circuit depth or qubit count:
\begin{equation}
    \tilde{\omega}_{i,j} = \omega_{i,j} + \lambda_k (1 - k_i \oplus k_j).
    \label{eq:omegatilde}
\end{equation}

\subsubsection{Road-Related Constraints}

Another class of constraints considered in this work addresses scenarios where certain pairs of nodes may not be directly connected, simulating real-world scenarios such as road closures or inaccessible paths.

To model this, we define a binary road constraint matrix $R_{i,j}$, where:
\begin{equation}
    R_{i,j} =
    \begin{cases}
        1 & \text{if node } i \text{ cannot be directly} \\& \text{connected to node } j, \\
        \\
        0 & \text{otherwise}.
    \end{cases}
\end{equation}

The corresponding penalty term in the Hamiltonian is given by:
\begin{equation}
\label{eq:road_constraint}
    R(x) = \lambda_R \sum_{\substack{i,j,t \\ i \neq j}} x_{i,t} \, x_{j,t+1} \, R_{i,j}.
\end{equation}

As this penalty term mirrors the structure of Equation~\ref{eq:capacity_constraints}, it can be seamlessly integrated into the cost matrix, thereby avoiding an increase in quantum circuit complexity:

\begin{equation}
    \tilde{\omega}_{i,j} = \omega_{i,j} + \lambda_R R_{i,j}.
\end{equation}

\subsubsection{Time-Step Constraints}

To incorporate time-related constraints into the TSP while remaining compatible with current quantum hardware limitations, we introduce a simplified variant inspired by the Vehicle Routing Problem with Time Windows (VRPTW)~\citep{cordeau2000vrp}. In VRPTW, certain nodes must be visited within predefined time windows due to operational or business requirements. However, a full VRPTW formulation requires a large number of additional variables, making it impractical for current quantum hardware~\citep{irie2019quantum}. To address this limitation, we propose a lightweight alternative that enforces simplified time constraints without increasing the qubit count. Specifically, we define \emph{time-step constraints}, which restrict specific cities to appear at predetermined positions in the tour.

We define a binary time constraint matrix $T_{i,t}$, where:
\begin{equation}
    T_{i,t} =
    \begin{cases}
        1 & \text{if city } i \text{ is not allowed}\\ &\text{to be visited at step } t, \\
        \\
        0 & \text{otherwise}.
    \end{cases}
\end{equation}

Based on this matrix, we introduce a penalty term in the Hamiltonian to discourage violations of the time constraints:
\begin{equation}
\label{eq:time_constraint}
    T(x) = \lambda_T \sum_{i,t} T_{i,t} x_{i,t},
\end{equation}

Consequently, the resulting cost function becomes:
\begin{equation}
    C(x) = D(x) + \tilde{P}(x) + T(x).
\end{equation}

In contrast to BNC and road constraints, time-step constraints cannot be embedded into the cost matrix and thus introduce a modest increase in circuit complexity.

\subsection{Grover-Inspired Mixer}
\label{subsec:grover_mixer}

In the TSP, many bitstrings represent invalid solutions, such as visiting multiple cities at the same time or revisiting a city. To improve efficiency, we employ a Grover-inspired mixer that limits mixing to partially feasible states, reducing the search space and increasing the likelihood of low-cost solutions. Constructing a superposition of fully feasible states is computationally intractable for large $n$, growing as $n!$, so we adopt a more practical initialization strategy.

The system of $n^2$ qubits is partitioned into $n$ registers of $n$ qubits each, denoted $|q_t\rangle$ for $t = 1, \dots, n$. Each register encodes the city visited in the step $t$ of the tour and must contain exactly one qubit in the $|1\rangle$ state, with the others in $|0\rangle$, corresponding to the Dicke state $|D_1^n\rangle$. Each register is initialized in an equal superposition of all valid one-hot encoded states. For example, when $n = 3$ (i.e., $3^2 = 9$ qubits), each register is initialized as:
\begin{align}
    U_s^t |0\rangle^{\otimes 3} = |D_1^3\rangle= \frac{1}{\sqrt{3}} \left( |100\rangle + |010\rangle + |001\rangle \right),
\end{align}
where $U_s^t$ is the unitary operator that prepares an equal superposition over all feasible states of the $t$-th register.

Following the structure of Equation~\ref{mixer_operator}, the mixer unitary for the $t$-th register in the circuit layer $i=\{1,\dots,p\}$ is defined as:
\begin{equation}
    U_m^t(\beta_i) =\ 
     U_s^t \left( I_d 
    - \left(1 - e^{-i\beta_i} \right) 
    |0\rangle\langle 0| \right) (U_s^t)^\dagger.
\end{equation}

Since each $U_m^t(\beta_i)$ acts on a distinct register, the overall mixer unitary applied to the full system is given by the tensor product:
\begin{equation}
    U_m(\beta_i) = \bigotimes_{t=1}^{n} U_m^t(\beta_i).
\end{equation}

By applying the mixer unitary defined in Equation~\ref{mixer_operator}, we restrict the quantum evolution to partially feasible subspaces. This ensures that each time step is associated with exactly one city visit, although some cities may still be revisited.
This approach reduces the size of the search space from $2^{\nu}$ with $\nu = n^2$ to $n^n$, representing a substantial reduction in search space compared to the standard $X$-Mixer.

Finally, the second term of Equation~\ref{eq:penalty_1} can be omitted when applying this mixer. As a result, the simplified cost function in Equation~\ref{eq:penalty_2} can be used instead.

\begin{figure}[ht!]
\centering
\begin{tikzpicture}
\node[scale=0.8] {
\begin{quantikz}
\lstick{$q_0^{\text{register }  0}$} & \gate[2]{U_s^{0}} & \qw & \gate[4]{U_c(\gamma)} & \qw & \gate[2]{U_m^{0}(\beta)} & \qw & \meter{} \\
\lstick{$q_1^{\text{register } 0}$} & \ghost{U_s^{0}}   & \qw & \ghost{U_c(\gamma)}   & \qw & \ghost{U_m^{0}(\beta)}   & \qw & \meter{} \\
\lstick{$q_2^{\text{register } 1}$} & \gate[2]{U_s^{1}} & \qw & \ghost{U_c(\gamma)}   & \qw & \gate[2]{U_m^{1}(\beta)} & \qw & \meter{} \\
\lstick{$q_3^{\text{register } 1}$} & \ghost{U_s^{1}}   & \qw & \ghost{U_c(\gamma)}   & \qw & \ghost{U_m^{1}(\beta)}   & \qw & \meter{} \\
\end{quantikz}
};
\end{tikzpicture}
\caption{Example of the QAOA circuit for $n=2$ cities and depth $p=1$. Each register is initialized via $U_s^t$, the cost operator $U_c(\gamma)$ is applied, and the Grover-inspired mixer $U_m^t(\beta)$ acts independently on each register to restrict evolution to partially feasible states. This guarantees that every time step corresponds to exactly one visited city, though city repetitions may still occur.}
\label{fig:QC_example}
\end{figure}
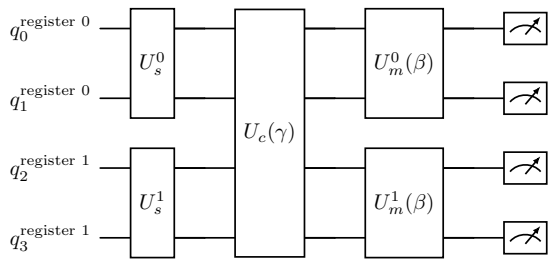

An example of the implemented quantum circuit is reported in Fig.~\ref{fig:QC_example}.

\subsection{Cl-QAOA for Large-Scale TSP}

The application of QAOA to large-scale TSP instances is constrained by the limited number of qubits available on current quantum hardware. To overcome this constraint, we adopt a hierarchical decomposition strategy: the original instance is partitioned into smaller sub-problems using classical machine learning (ML)-based clustering, each sub-problem is solved with QAOA, and the partial solutions are combined into a global tour.

\begin{figure*}[ht!]
    \centering
    \includegraphics[width=1\textwidth]{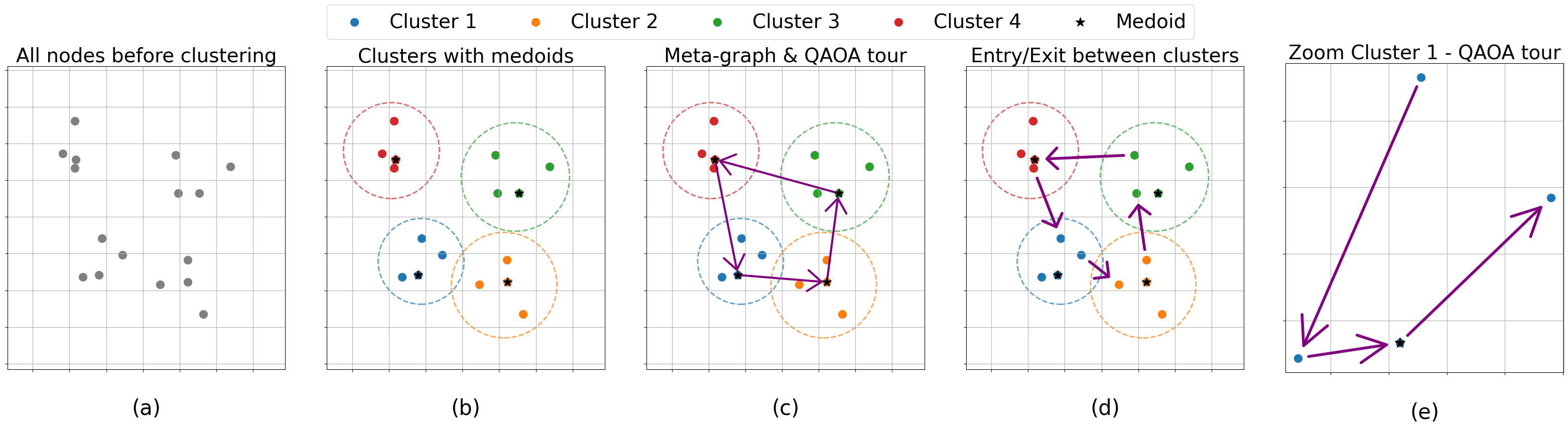}
    \caption{Overview of the clustering-based QAOA pipeline for the TSP.
(a) Spatial distribution of cities before clustering.
(b) Problem decomposition via clustering into sub-problems of manageable size.
(c) Meta-TSP resolution on the cluster medoids.
(d) Selection of entry and exit nodes between consecutive clusters.
(e) Solution of each sub-TSP using QAOA within a cluster.}
    \label{example_QAOA+clustering}
\end{figure*}

We refer to this approach as clustering QAOA (Cl-QAOA), and it enables scalable quantum optimization through recursive decomposition. The overall workflow is summarized in Algorithm~\ref{alg:cl-qaoa}, and a visual representation is provided in Figure~\ref{example_QAOA+clustering}.
\begin{algorithm}
\caption{clustering QAOA (Cl-QAOA) Workflow}
\label{alg:cl-qaoa}
\begin{enumerate}
    \item \textbf{Cluster cities:} Partition the TSP instance into $k \le N_{\max}$ clusters using a chosen clustering method, where $N_{\max}$ denotes the maximum problem size solvable by QAOA.
    \item \textbf{Construct meta-graph:} Build a graph whose nodes represent clusters, where each cluster node corresponds to the medoid of the cluster. Edge weights are defined as the inter-cluster distances.
    \item \textbf{Solve meta-TSP:} Apply QAOA to the meta-graph to determine the optimal visiting order of clusters.
    \item \textbf{Select entry/exit nodes:} For each consecutive cluster pair, choose the nodes that minimize transition costs.
    \item \textbf{Solve sub-TSPs:} For each cluster, whose size is denoted by $m$:
    \begin{itemize}
        \item If $m \le N_{\max}$, solve the sub-problem with QAOA (respecting entry/exit constraints).
        \item Otherwise, recursively apply Steps~1–5.
    \end{itemize}
    \item \textbf{Assemble global tour:} Concatenate local tours according to the meta-order and compute the total cost.
\end{enumerate}
\end{algorithm}

Several clustering techniques can be employed for decomposition, each with specific trade-offs  in terms of scalability, interpretability, and robustness. In this work, we adopt \emph{Agglomerative Hierarchical Clustering (AHC)} due to its robustness against noise and outliers, compatibility with distance matrices, and natural alignment with hierarchical decomposition~\citep{zepeda2013hierarchical}. AHC offers a good balance between interpretability, flexibility, and clustering quality, making it suitable for the TSP instances considered.

To benchmark Cl-QAOA, we compare against two widely adopted metaheuristics for large-scale TSP: Simulated Annealing (SA)~\citep{kirkpatrick1983optimization} and Ant Colony Optimization (ACO)~\citep{dorigo2004ant}. Both methods follow standard formulations with minor enhancements for robustness and convergence. Hyperparameters are tuned empirically to balance solution quality and runtime. SA employs a logarithmic cooling schedule with reheating and a final 2-opt refinement, while ACO uses adaptive pheromone bounds and heuristic weighting. Detailed algorithmic settings are provided in Appendix~\ref{appendix:classical}.

\subsection{Time Modeling for QAOA and Cl-QAOA}
\label{subsec:time_model}
In this subsection, we formulate a computational time model for both the standard QAOA applied to the TSP and its clustering-based variant, Cl-QAOA. The objective is to characterize their scaling behavior on real quantum hardware, identifying the main performance bottlenecks and advantages of each approach for large problem instances.

\subsubsection{QAOA Runtime Model for TSP}

The runtime of QAOA on quantum hardware can be decomposed into two main contributions: the optimization phase and the final sampling stage. Each iteration of the classical optimization loop involves executing the parameterized circuit $s$ times to estimate the expectation value of the cost Hamiltonian, followed by parameter updates. After convergence, an additional set of $s^\ast$ executions is performed to sample candidate solutions. Neglecting constant overheads, the total wall-clock time can be expressed as:
\begin{equation}
    T_{\text{QAOA}} \approx 
    \underbrace{I \cdot s \cdot t_{\text{shot}}}_{T_{\text{QAOA}}^{\text{optimization}}}
    + \underbrace{s^\ast \cdot t_{\text{shot}}}_{T_{\text{QAOA}}^{\text{final sampling}}},
    \label{eq:QAOA_hardware_time}
\end{equation}
where $I$ is the number of classical iterations and $t_{\text{shot}}$ denotes the time per circuit execution, which scales with the number of two-qubit gates. For dense QUBO encodings of TSP, the cost layer requires $\mathcal{O}(n^4)$ two-qubit interactions, resulting in a circuit depth of $\mathcal{O}(n^4 \cdot p)$ for $p$ QAOA layers \citep{weidenfeller2022scaling,he2024quantum}. The execution time per shot therefore scales as
\[
t_{\text{shot}} = \mathcal{O}(n^4 \cdot p)\,\tau_{\text{CNOT}},
\]
where $\tau_{\mathrm{CNOT}}$ is the duration (time) of a two-qubit (CNOT) gate. This relation forms the basis for analyzing how $I$, $s$, $s^\ast$, and $p$ affect the overall complexity.

Routing overhead can further increase depth by a constant factor or more.
Substituting $t_{\text{shot}}$ into Equation~\ref{eq:QAOA_hardware_time} and neglecting lower-order terms yields:
\begin{equation}
\begin{split}
    &T_{\text{QAOA}}^{\text{optimization}} = \mathcal{O}(I \cdot s \cdot n^4 \cdot p), \\
    &T_{\text{QAOA}}^{\text{final sampling}} = \mathcal{O}(s^\ast \cdot n^4 \cdot p).
    \label{eq:time_QAOA}
\end{split}    
\end{equation}

Empirical studies suggest that $I$ increases with $p$ due to the enlarged parameter space and more complex energy landscapes \citep{weidenfeller2022scaling}. Characterizing this dependence is essential for predicting overall runtime and assessing the feasibility of QAOA for large-scale TSP.

The number of shots $s$ required to estimate the expectation value of the cost Hamiltonian within relative error $\delta$ satisfies:
\begin{align}
\sqrt{\frac{\mathrm{Var}(H_C)}{s}} \le \delta \, \langle H_C\rangle, \qquad
s \ge \frac{\mathrm{Var}(H_C)}{\delta^2 \langle H_C\rangle^2}.
\end{align}
For the QUBO encoding of TSP, $\langle H_C\rangle = \mathcal{O}(n^4)$ and $\mathrm{Var}(H_C) = \mathcal{O}(n^8)$ (see Appendix~\ref{appendix:scaling}), yielding:
\[
s = \mathcal{O}\!\left(\frac{1}{\delta^2}\right).
\]
This term contributes multiplicatively to the optimization runtime in Equation~\ref{eq:time_QAOA}.

The number of shots $s^\ast$ required for the final sampling stage depends on the probability $\rho$ of drawing a solution with cost $\leq \alpha C_{\text{opt}}$, $\alpha\geq 1$. For a target success probability $\pi$:
\[
s^\ast \ge \frac{\log(1-\pi)}{\log(1-\rho)}.
\]
Under a uniform distribution, $\rho = M(\alpha)/2^\nu$, where $M(\alpha)$ is the number of states satisfying the cost bound and $\nu = n^2$ is the number of qubits. For $\alpha = 1$, $M(1)\approx 1$, yielding $\rho \approx 2^{-\nu}$ and an exponential scaling:
\[
s^\ast \approx 2^\nu \log\!\left(\frac{1}{1-\pi}\right).
\]
However, in practical scenarios, $M(\alpha)$ grows rapidly with $\alpha$ and depends on the cost landscape, so $\rho$ can be orders of magnitude larger than $2^{-\nu}$. Consequently, the scaling may shift from exponential to sub-exponential or even polynomial, particularly when the optimized state concentrates probability on low-cost regions.

Circuit depth $p$ critically impacts runtime and solution quality, yet its scaling with problem size $n$ remains largely empirical. Since $p$ influences both the optimization loop and the final sampling cost (Equation~\ref{eq:time_QAOA}), understanding this dependence is essential. Increasing $p$ can improve solution quality and boost $\rho$, reducing $s^\ast$, but at the expense of higher gate counts and longer execution times. Characterizing this trade-off is key to assessing the feasibility of QAOA for TSP on near-term hardware (see Subsection~\ref{subsec:time_QAOA}).

\subsubsection{Cl-QAOA Runtime Model}
The runtime of Cl-QAOA depends on the number of QAOA instances executed across two contexts: meta-TSP problems ordering clusters at each recursion level and sub-TSP problems within clusters of size at most $N_{\max}$. For an original instance of size $n$, the algorithm recursively partitions clusters exceeding $N_{\max}$ until all clusters satisfy this bound. Let $L$ denote the recursion depth. In the worst case, $L = \mathcal{O}(n)$, while balanced partitions yield $L = \mathcal{O}(\log_{N_{\max}} n)$. The total number of QAOA calls satisfies:
\[
N_Q = \mathcal{O}\!\left(\frac{n}{N_{\max}}\right),
\]
since final clusters dominate over meta-TSP overhead. Each QAOA instance processes at most $N_{\max}$ nodes, so the runtime is:
\begin{equation}
    T_{\mathrm{Cl\text{-}QAOA}}(n)
    = \mathcal{O}\!\left(\frac{n \cdot T_{\mathrm{QAOA}}(N_{\max})}{N_{\max}}\right).
    \label{eq:time_clqaoa}
\end{equation}

% \begin{equation}
%     T_{\mathrm{Cl\text{-}QAOA}}^{\mathrm{seq}}(n) =
%     \mathcal{O}\!\left(\frac{n \cdot T_{\mathrm{QAOA}}(N_\max)}{N_\max}\right).
%     \label{eq:time_clqaoa}
% \end{equation}
This bound holds regardless of cluster distribution, as each node is solved exactly once and meta-TSP overhead grows at most linearly in the worst case.

\begin{comment}
The runtime of Cl-QAOA should exhibit linear scaling with the problem size $n$.
In particular, we have modelled the clustering decomposition as a $N_{\text{max}}$-ary tree, where $N_{\text{max}}$ is the maximum problem size solvable by QAOA at the base level. A problem of size $m$ is split into $N_{\text{max}}$ sub-problems of size approximately $m/N_{\text{max}}$, and the recursion stops when $m \le N_{\text{max}}$, yielding a tree depth
\begin{equation}
   L = \mathcal{O}\left(\log_{N_{\text{max}}}(n)\right). 
\end{equation}

The total number of QAOA evaluations is

\begin{align}
    N_Q &= 1 + N_{\text{max}} + N_{\text{max}}^2 + \dots + N_{\text{max}}^{L-1} \notag
    \\ &= \frac{N_{\text{max}}^{L} - 1}{N_{\text{max}} - 1}
    \approx \frac{n}{N_{\text{max}} - 1},
\end{align}

using $N_{\text{max}}^{L} \approx n$. Since each call processes at most $N_{\text{max}}$ nodes, the sequential runtime is
\begin{equation}
    T_{\mathrm{H\text{-}QAOA}}^{\mathrm{seq}}(n)
    \simeq \frac{n}{N_{\text{max}} - 1} \, T_{\mathrm{QAOA}}(k),
\end{equation}

where $T_{\mathrm{QAOA}}(N_{\text{max}})$ is the time taken by QAOA run of size $N_{\text{max}}$. Under full parallelization across levels, the runtime becomes
\begin{equation}
    T_{\mathrm{H\text{-}QAOA}}^{\mathrm{par}}(n)
    \simeq \log_{N_{\text{max}}}(n)\, T_{\mathrm{QAOA}}(N_{\text{max}}).
\end{equation}

\end{comment}

\section{Experimental Setup and Protocols}
\label{sec:experimental_setup}
This section presents the experimental setup and protocols employed in our analysis. Specifically, we describe the datasets used to evaluate the TSP under both the QAOA and Cl-QAOA frameworks, outline the computational environment, define the metrics used to assess solution quality, and detail the algorithmic configurations and experimental procedures.
Additionally, we outline the algorithmic structure of QAOA combined with ML techniques employed to address large-scale problem instances.

\subsection{Datasets}

To evaluate the performance of QAOA, we consider both synthetic and realistic TSP datasets. This subsection provides a detailed description of the two datasets used in our experiments.

\subsubsection{Synthetic Dataset}

The first dataset employed is synthetic, generated by sampling from a random distribution. Specifically, the cost matrix is constructed as an $n \times n$ matrix with entries randomly sampled from the interval $[0, 10]$. Diagonal elements are set to zero, reflecting the zero cost of remaining at the same node. The matrix is intentionally asymmetric to simulate unbalanced travel costs.

This dataset serves as a benchmark to assess algorithmic performance in the absence of inherent structure. Since the data lacks natural ordering or spatial patterns, identifying optimal solutions is generally more challenging compared to structured or realistic datasets.

\subsubsection{Realistic Dataset: Milan Subway Network}

To assess algorithm performance in a real-world setting, we employed a dataset derived from the geographic locations of Milan's subway stations, obtained from open-source GTFS data\footnote{\url{https://dati.comune.milano.it/gtfs.zip}}.
From this dataset, we extracted the coordinates of the most frequently used stations.
A realistic cost matrix was then constructed using OpenStreetMap (OSM) data\footnote{Computed with the \texttt{osmnx} Python library: \url{https://github.com/gboeing/osmnx}}.

Rather than relying on Euclidean distances, we constructed a road network graph that accounts for real-world constraints such as intersections, road segments, and one-way streets. Pairwise distances between stations were computed using Dijkstra's algorithm~\citep{dijkstra1959}, ensuring that computed paths reflect realistic travel routes and accessibility. To better approximate travel conditions, the distance-based matrix was converted into a time-based one by dividing each edge length by the maximum allowed speed from OSM metadata. The resulting matrix is asymmetric, capturing variability in travel times due to road characteristics.

This dataset supports practical applications such as:
\begin{itemize}
    \item \textbf{Inspection and Maintenance Planning:} Prioritizing interventions on critical or high-traffic stations.
    \item \textbf{Network Vulnerability Analysis:} Identifying shortest paths between major nodes for emergency planning and detour strategies.
\end{itemize}

\begin{figure}[ht!]
    \centering
    \includegraphics[width=0.9\columnwidth]{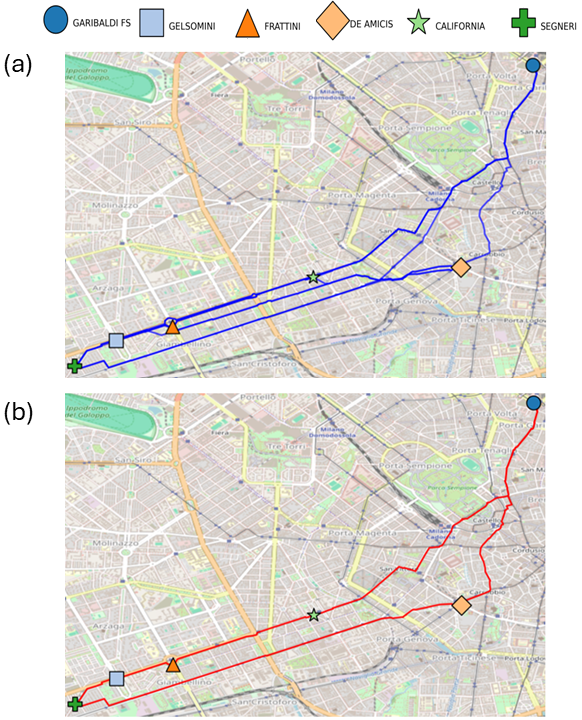}
    \caption{Example of the Milan subway dataset used for TSP experiments. Panel (a) shows all station connections in blue, representing the network graph used for cost computation. Panel (b) highlights in red the solution of the TSP problem over the same network.}
    \label{OSM_example}
\end{figure}

An example of the dataset employed is shown in Fig.~\ref{OSM_example}.

To enable controlled scalability testing, for each fixed problem size $n$ we constructed the dataset as follows. First, stations were ranked according to their traffic frequency, as inferred from the GTFS data, selecting those that accommodate the highest number of trains. Then, to introduce variability while keeping the problem size fixed, we generated multiple instances of size $n$ by randomly sampling from the top $2n$ stations. This procedure ensures that each subset maintains high relevance in terms of traffic while providing diverse configurations for benchmarking algorithm performance.

\subsection{Computational Environment}

All simulations were performed using the \texttt{Qiskit} framework. All simulations for problem sizes up to $n = 5$ were executed on a local high-performance workstation featuring $96$ CPU cores. The computational capabilities of this system significantly reduced simulation time, enabling the execution of a large number of distinct runs within a feasible timeframe.

As previously discussed, the QUBO formulation of the TSP requires $n^2$ qubits, where $n$ is the number of cities/nodes in the problem instance. Simulating such a quantum system requires storing the full quantum state vector, which consists of $2^{\nu}$, with $\nu = n^2$, complex amplitudes. Assuming double-precision floating-point representation, the memory requirements grow exponentially with $n$, as shown in Table~\ref{tab:memory}.
\begin{table}[!ht]
    \centering
    \begin{tabular*}{\linewidth}{@{\extracolsep{\fill}}ccc}
        \toprule
        \# Cities & \# Qubits & RAM Required [GiB] \\
        \midrule
        4 & 16 & $9.8 \times 10^{-4}$ \\
        5 & 25 & $0.5$ \\
        6 & 36 & $1024$ \\
        \bottomrule
    \end{tabular*}
    \caption{Estimated memory requirements for storing QUBO-based TSP instances. Values are computed assuming double-precision complex amplitudes.}
    \label{tab:memory}
\end{table}
Local simulations are feasible only for instances up to $n = 5$. For larger instances, the exponential growth in memory requirements imposes the use of more powerful high-performance computing (HPC) resources.

\paragraph{CINECA Infrastructure}

To simulate larger TSP instances, we leveraged the Leonardo supercomputer hosted by the CINECA consortium. Leonardo is a pre-exascale system designed for large-scale scientific computing and AI workloads. For our experiments, we simulated the case $n = 6$, corresponding to a 36-qubit quantum system. The simulation employed 8 compute nodes, each equipped with 4 NVIDIA A100 GPUs (64~GB per GPU) and a single CPU. The statevector was distributed across all 32 GPUs using the \texttt{cusvaer} backend from \texttt{Qiskit}, integrated with NVIDIA's cuQuantum Appliance~\citep{bayraktar2023cuquantum}. This backend partitions the statevector across multiple GPUs and nodes, enabling large-scale simulation with GPU acceleration. Each GPU allocated up to 32~GB for storing the statevector, reserving the remaining memory for inter-GPU communication and auxiliary operations essential for distributed execution.

All quantum circuits were transpiled with \texttt{Qiskit} at optimization level~3 to reduce gate count and circuit depth, thereby improving simulation efficiency.

\subsection{Performance Evaluation Metrics}

We evaluate solution quality using two normalized approximation ratios ~\citep{choi2019ar}, both bounded in $[0,1]$:
\begin{align}
    AR_{\text{exp}} &= \frac{\langle C \rangle - C_{\text{worst}}}{C_{\text{opt}} - C_{\text{worst}}}, \\
    AR_{\text{min}} &= \frac{C_{\text{min}} - C_{\text{worst}}}{C_{\text{opt}} - C_{\text{worst}}},
\end{align}
where $\langle C \rangle$ is the expected cost over sampled solutions and $C_{\text{min}}$ is the minimum cost observed.  
Thus, $AR_{\text{exp}}$ captures the expected quality of a sampled solution, while $AR_{\text{min}}$ reflects the quality of the best solution found.  
The quantities $C_{\text{opt}}$ and $C_{\text{worst}}$ are obtained via exhaustive enumeration of all $n^2$-bit configurations.  
Under this normalization, $AR_\text{exp} = 1$ and $AR_\text{min} = 1$ correspond to the optimal solution, while $AR_\text{exp} = 0$ and $AR_\text{min} = 0$ refer to the worst-case configuration.

For large instances, computing $C_{\text{opt}}$ is infeasible due to the exponential growth of the solution space. In these cases, we adopt a pragmatic benchmarking strategy using ACO as a reference, given its strong empirical performance on TSP. We define:
\begin{align}
    \widetilde{AR}_{\text{Cl-QAOA}} &= \frac{C_{\text{Cl-QAOA}}}{C_{\text{ACO}}}, \\
    \widetilde{AR}_{\text{SA}} &= \frac{C_{\text{SA}}}{C_{\text{ACO}}}, 
\end{align}
where $C_{\text{Cl-QAOA}}$, $C_{\text{ACO}}$, and $C_{\text{SA}}$ denote the minimum costs obtained by applying Cl-QAOA, ACO, and SA respectively.  
Because all ratios are normalized by $C_{\text{ACO}}$, the approximation ratio is directly aligned with solution quality: the smaller the value, the better the solution. In particular, $\widetilde{AR} = 1$ indicates performance equal to ACO, $\widetilde{AR} > 1$ corresponds to worse performance, and $\widetilde{AR} < 1$ denotes a solution strictly better than that found by ACO. This makes the comparison across methods both intuitive and consistent.

\subsection{Protocols for QAOA and Cl-QAOA}
\label{subsec:hyperparam_sampling}

In this subsection, we describe the experimental protocols adopted for QAOA and its hierarchical extension Cl-QAOA, including optimizer configuration, hyperparameter initialization, shot allocation, and circuit-depth analysis. These choices aim to balance solution quality, computational efficiency, and scalability across different problem sizes.

\subsubsection{QAOA Configuration}

Following prior studies~\citep{campbell2022qaoa,turati2022qaoa}, we adopt the \texttt{COBYLA} optimizer~\citep{powell1994cobyla}, which offers a favorable trade-off between solution quality and computational efficiency~\citep{singh2023optimizers,fernandezpendas2022optimizers}.

The initial QAOA parameters $\gamma_i$ and $\beta_i$ for $i = 1, \dots, p$ are randomly sampled from $[0, 2\pi]$ and refined by the classical optimizer, with a maximum of $200$ iterations. This iteration limit was sufficient to reach convergence in all tested cases.

Penalty coefficients for canonical and logistical constraints ($\lambda_p$, $\lambda_k$, $\lambda_R$, and $\lambda_T$ in Equations~\ref{eq:penalty_2}, \ref{eq:capacity_constraints}, \ref{eq:road_constraint}, and \ref{eq:time_constraint}) are set as:
\[
\lambda = \max_{i,j}(\omega_{i,j}) \cdot n.
\]
This choice follows a common strategy in QUBO formulations, where penalty weights are set to ensure that any feasible solution has a lower cost than any infeasible one~\cite{glover2022qubo, lucas2014ising, ayodele2022penalty}.
Specifically, using a value proportional to the maximum edge weight and the problem size provides a conservative bound, guaranteeing that any constraint violation incurs a higher cost than the worst-case feasible solution.
Alternative approaches found in the literature include setting penalties based on the full range of the objective function or tuning them empirically~\cite{ayodele2022penalty}. However, excessively large penalties can distort the energy landscape, potentially degrading optimization performance ~\cite{ayodele2022penalty,mirkarimi2024quantum}.
Selecting appropriate penalty values remains a challenging task and may be better addressed in future work through adaptive or data-driven strategies.

\subsubsection{QAOA Shot Analysis}

This analysis is motivated by two considerations. First, evaluating performance at minimal circuit depth provides insight into the algorithm's ability to identify high-quality solutions under constrained quantum resources, a regime particularly relevant for near-term hardware. Second, varying the number of shots $s$ exposes the trade-off between sampling accuracy and computational cost: increasing $s$ improves the estimation of output probabilities but also increases the runtime, making it essential to determine the minimal shot count that yields reliable results. These aspects are directly determined by how QAOA solutions are obtained from repeated circuit executions.

In this context, a \emph{shot} corresponds to a single measurement of all qubits after running the quantum circuit, producing one bitstring sampled from its output distribution. Aggregating multiple shots allows us to approximate the probability of each bitstring and, consequently, to assess the quality of the best solution observed.

All experiments in this section were performed at fixed QAOA depth $p=1$, varying the number of shots $s$ to quantify the impact of finite sampling. For small problem sizes, we tested
\[
s \in \{10,\,100,\,500,\,1000,\,2000,\,5000\},
\]
while for the largest instance considered ($n=6$), practical constraints restricted the sweep to $s \in \{500,\,2000\}$. Each configuration (dataset, constraint, $s$) was repeated multiple times using different random seeds and independently sampled cost matrices; the exact repetition counts are reported in Tables~\ref{tab:shot_analysis_combined} and \ref{tab:shot_n6} in the Results section.

For each run, we computed the minimum approximation ratio $AR_{\text{min}}$ by selecting, among all sampled bitstrings across all iterations, the tour with the lowest cost.

\subsubsection{QAOA Circuit-Depth Analysis}

Depth-dependent experiments were performed at fixed sampling budgets to assess how increasing $p$ affects the expectation-based approximation ratio $AR_{\text{exp}}$ and runtime. The tested ranges were:
\begin{itemize}
    \item $p=1,\dots,10$ for $n=3,4$;
    \item $p=1,\dots,4$ for $n=5$;
    \item $p=1,\dots,3$ for $n=6$.
\end{itemize}
For $n=3,4,5$ the analysis used $s=500$ shots per classical iteration; while $s=2000$ shots for $n=6$. 

This analysis aims to determine whether increasing the circuit depth leads to a significant improvement in solution quality.
While deeper circuits enhance the expressivity of the variational ansatz, they also increase computational complexity and susceptibility to noise and decoherence, which are critical factors for near-term quantum hardware.

\subsubsection{Cl-QAOA Analysis}  

For large instances ($n>6$), the Cl-QAOA algorithm was applied, varying the maximum sub-problem size solvable by QAOA ($N_{\text{max}}$) to evaluate scalability. For each problem size $n$, $N_{\text{max}}$ was swept from $3$ to $20$. Instances with $N_{\text{max}} = 3,4,5$ were solved on a local workstation, while $N_{\text{max}} = 6$ instances were executed on the CINECA infrastructure.  

For $N_{\text{max}}$ between $7$ and $20$, QAOA was replaced by an exact brute-force solver due to memory limitations, and the resulting metrics $\widetilde{AR}_{\text{Cl-QAOA}}$ were reported as ``Estimated $\widetilde{AR}_{\text{Cl-QAOA}}$'', providing an upper bound on the achievable performance of the Cl-QAOA approach.  

Both synthetic and OSM Milan subway datasets were considered without additional logistic constraints, with problem sizes $n = 20, 40, 80, 130$ and multiple instances per size corresponding to different cost matrices. Solution quality and computational time were measured to assess scalability.

\section{Results}
\label{sec:results}

This section presents the experimental results obtained using the QAOA algorithm on small-scale datasets (up to $n = 6$ cities), as well as the performance of the clustering QAOA-based TSP pipeline on larger instances (up to $n = 130$ cities).

\subsection{Shot Analysis}
\label{subsec:shot_analysis}

Tables~\ref{tab:shot_n3}--\ref{tab:shot_n5} report $AR_{\text{min}}$ for each configuration, along with the number of independent runs (``Runs''), which correspond to problem instances generated under identical constraints and hyperparameters but with different cost matrices.

The algorithm consistently identifies optimal solutions even with a relatively small number of shots. Specifically, $AR_{\text{min}} = 1$ is achieved across all constraint configurations with as few as 10 shots for $n = 3$, 100 shots for $n = 4$, and 500 shots for $n = 5$.

%%%%%%%%%%%%%%%%%%%%%%%%%%%%%%
\paragraph{Results for n = 6 from Simulations on the Leonardo HPC Infrastructure}

The corresponding results are reported in Table~\ref{tab:shot_n6}.

For the instance with $n = 6$, a shot count of 2000 was sufficient to sample the optimal solution during execution, yielding $\langle AR_{\text{min}} \rangle = 1$. This result indicates that high-quality solutions can be obtained with a relatively modest sampling effort.

\subsection{Circuit Depth Analysis}
\label{subsec:depth_analysis}

Results for solution quality are shown in Figures~\ref{fig:all_n4_n5_n6_AR}(a) and (b).

Results indicate that increasing $p$ generally improves $AR_{\text{exp}}$, particularly for smaller instances ($n = 3$ and $n = 4$), where modest depth increases yield noticeable gains. For $n = 5$, improvements are less pronounced, likely due to the exponential growth of the solution space. Nonetheless, the trend confirms that deeper circuits enhance average solution quality, even when optimal solutions are already accessible at shallow depths, as shown in the shot analysis.

\paragraph{Results for n = 6 from Simulations on the Leonardo HPC Infrastructure}

We extended the depth analysis to TSP instances with $n=6$ evaluating the approximation ratio of the expectation value, $AR_{\text{exp}}$ using the Leonardo high-performance computing infrastructure as presented in Figure~\ref{fig:all_n4_n5_n6_AR}.

The results for $n=6$ follow the same trend observed for $n=5$, suggesting that larger search spaces require higher circuit depths to achieve noticeable improvements. Nevertheless, even at low values of $p$, the algorithm consistently delivers solutions of good quality.

\begin{table*}[!ht]
    \centering
    \begin{subtable}[!ht]{0.45\textwidth}
        \centering
        \caption{$n = 3$}
        \resizebox{\textwidth}{!}{
        \begin{tabular}{cccc|c}
            \toprule
            Dataset & Constraint & Shots & Runs & $\langle AR_{\text{min}} \rangle$ \\
            \midrule
            Synthetic & - & 10-5000 & 5 & 1.000 \\
            OSM Milan & - & 10-5000 & 5 & 1.000 \\
            \hline
            Synthetic & BNC & 10-5000 & 15 & 1.000 \\
            OSM Milan & BNC & 10-5000 & 15 & 1.000 \\
            \hline
            Synthetic & road & 10-5000 & 25 & 1.000 \\
            \hline
            Synthetic & time & 10-5000 & 15 & 1.000 \\
            OSM Milan & time & 10-5000 & 15 & 1.000 \\
            \bottomrule
        \end{tabular}
        }
        \label{tab:shot_n3}
    \end{subtable}
    \hfill
    \begin{subtable}[!ht]{0.45\textwidth}
        \centering
        \caption{$n = 4$}
        \resizebox{\textwidth}{!}{
        \begin{tabular}{cccc|c}
            \toprule
            Dataset & Constraint & Shots & Runs & $\langle AR_{\text{min}} \rangle$ \\
            \midrule
            Synthetic & - & 10-5000 & 5 & 1.000 \\
            OSM Milan & - & 10-5000 & 5 & 1.000 \\
            \hline
            Synthetic & BNC & 10-5000 & 15 & 1.000 \\
            OSM Milan & BNC & 10-5000 & 15 & 1.000 \\
            \hline
            Synthetic & road & 10-5000 & 25 & 1.000 \\
            \hline
            Synthetic & time & 10 & 15 & 0.998 \\
            Synthetic & time & 100-5000 & 15 & 1.000 \\
            OSM Milan & time & 10-5000 & 15 & 1.000 \\
            \bottomrule
        \end{tabular}
        }
        \label{tab:shot_n4}
    \end{subtable}
    \\
    \begin{subtable}[!ht]{0.45\textwidth}
        \centering
        \caption{$n = 5$}
        \resizebox{\textwidth}{!}{
        \begin{tabular}{cccc|c}
            \toprule
            Dataset & Constraint & Shots & Runs & $\langle AR_{\text{min}} \rangle$ \\
            \midrule
            Synthetic & - & 10 & 5 & 0.999 \\
            Synthetic & - & 100-5000 & 5 & 1.000 \\
            OSM Milan & - & 10 & 5 & 0.999 \\
            OSM Milan & - & 100-5000 & 5 & 1.000 \\
            \hline
            Synthetic & BNC & 10 & 20 & 0.987 \\
            Synthetic & BNC & 100-5000 & 20 & 1.000 \\
            OSM Milan & BNC & 10 & 20 & 0.992 \\
            OSM Milan & BNC & 100-5000 & 20 & 1.000 \\
            \hline
            Synthetic & road & 10 & 25 & 0.998 \\
            Synthetic & road & 100-5000 & 25 & 1.000 \\
            \hline
            Synthetic & time & 10 & 25 & 0.996 \\
            Synthetic & time & 100 & 25 & 0.999 \\
            Synthetic & time & 500-5000 & 25 & 1.000 \\
            OSM Milan & time & 10-100 & 25 & 0.999 \\
            OSM Milan & time & 500-5000 & 25 & 1.000 \\
            \bottomrule
        \end{tabular}
        }
        \label{tab:shot_n5}
    \end{subtable}
    \vspace{0.5cm}
    \caption{Shot analysis for $n = 3$ (a), $n = 4$ (b) and $n = 5$ (c) with $p = 1$. Each entry in the ``Shots'' column represents a range of shot counts (e.g., 10-5000) over which the corresponding data have been averaged. The column $\langle AR_{\text{min}} \rangle$ reports the mean value of the approximation ratio $AR_{\text{min}}$ across all runs within each shot range.}
\label{tab:shot_analysis_combined}
\end{table*}

\begin{table*}[!ht]
    \centering
    % {\large\textbf{Data obtained from simulations executed on the Leonardo HPC infrastructure.}}
    \begin{center}
        {\textbf{$n=6$}}
    \end{center}   
    \resizebox{0.5\textwidth}{!}{
    \begin{tabular}{cccc|c}
        \toprule
        Dataset & Constraint & Shots & Runs & $\langle AR_{\text{min}} \rangle$ \\
        \midrule
        Synthetic & - & 500 & 1 & 0.999 \\
        Synthetic & - & 2000 & 1 & 1.000 \\
        OSM Milan & - & 500-2000 & 1 & 1.000 \\
        \hline
        Synthetic & BNC & 500-2000 & 1 & 1.000 \\
        OSM Milan & BNC & 500 & 1 & 0.999 \\
        OSM Milan & BNC & 2000 & 1 & 1.000 \\
        \hline
        Synthetic & road & 500-2000 & 1 & 1.000 \\
        \hline
        Synthetic & time & 500-2000 & 1 & 1.000 \\
        OSM Milan & time & 500-2000 & 1 & 1.000 \\
        \bottomrule
    \end{tabular}
    }
    \vspace{0.5cm}
    \caption{Shot analysis for $n = 6$ and $p = 1$. Each entry in the ``Shots'' column represents a range of shot counts (e.g., 500-2000) over which the corresponding data have been averaged. The column $\langle AR_{\text{min}} \rangle$ reports the mean value of the approximation ratio $AR_{\text{min}}$ across all runs within each shot range.}
    \label{tab:shot_n6}
\end{table*}

\FloatBarrier
\begin{strip}
    \centering
    \includegraphics[width=0.78\textwidth]{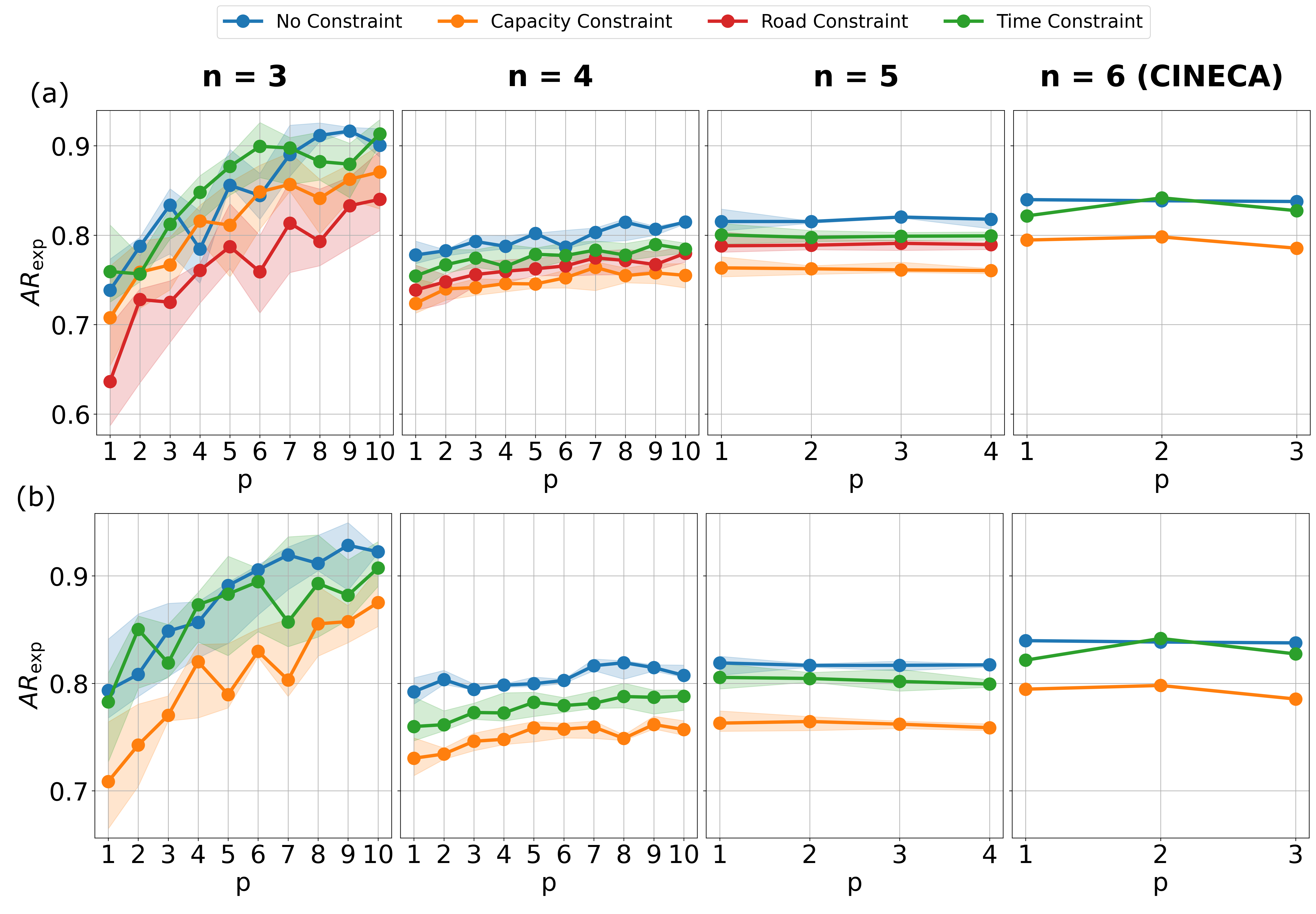}
    \captionof{figure}{QAOA performance across both the random synthetic (a) and the realistic datasets derived from the Milan subway system (b) for $n = 3$, $n = 4$, $n = 5$ and $n=6$ using the CINECA infrastructure. Each plot displays the approximation ratio of the expectation value, $AR_{\text{exp}}$ as a function of the circuit depth $p$.}
    \label{fig:all_n4_n5_n6_AR}
\end{strip}

\subsection{Time Complexity of QAOA}
\label{subsec:time_QAOA}

\begin{figure}[!ht]
    \centering
    \includegraphics[width=1\columnwidth]{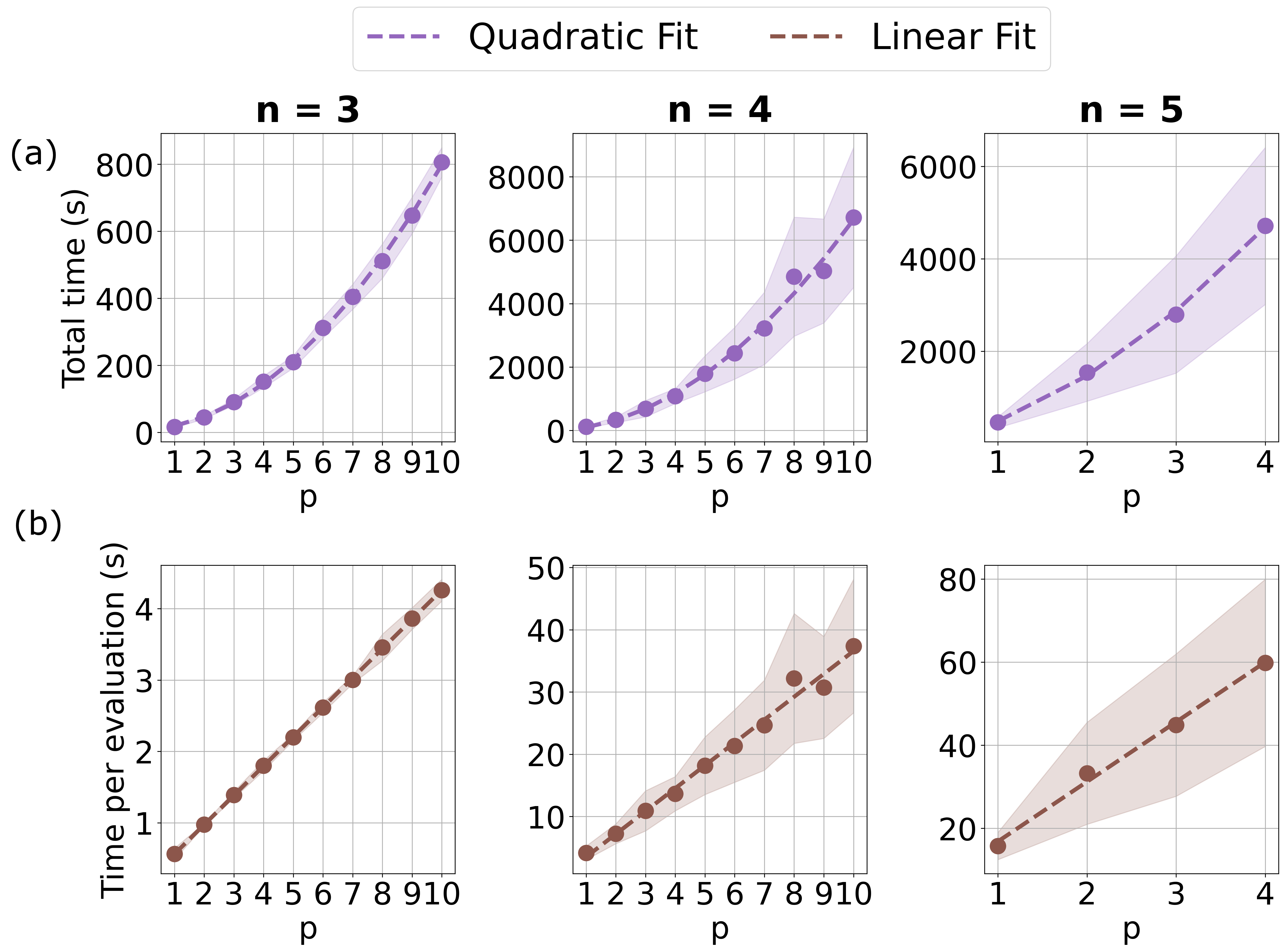}
\caption{(a) Total QAOA runtime as a function of circuit depth, with a quadratic fit. 
(b) Runtime of a single circuit evaluation, with a linear fit. The data are obtained from analyses performed on the realistic Milan metro dataset.}
    \label{fig:time_QAOA_vs_p}
\end{figure}

Figure~\ref{fig:time_QAOA_vs_p} reports the computational time as a function of the circuit depth. Panel~(a) shows the full QAOA runtime, including the classical optimization loop, whereas panel~(b) isolates the cost of a single circuit execution without classical optimization.

The linear trend observed in panel~(b) shows that increasing circuit depth results in a proportional growth in runtime. When the classical optimization is performed, as described in Subsection~\ref{subsec:time_model}, the scaling becomes quadratic in $p$, consistent with the relation $I \propto p$. 

Moreover, we investigated the scaling of QAOA with increasing problem size by estimating the critical depth $p^\ast(n)$ required to achieve a target approximation ratio $AR_{\text{exp}}^{\text{target}}=0.95$. 
We modeled the dependence of the approximation ratio on the circuit depth using an exponential saturation model
\[
AR_{\text{exp}}(p) = 1 - A e^{-k p},
\]
where $A$ and $k$ are fitting parameters. 
This choice is motivated both by theory \citep{farhi2014qaoa}, which predicts that QAOA approaches the ground state at large $p$, and by the empirical model proposed in \citep{zhou2020quantum}, implying that $AR_{\text{exp}}(p)$ grows monotonically with $p$ and asymptotically approaches 1.

Figure~\ref{fig:pvsn}(a) shows the measured approximation ratio as a function of $p$ in the tested range, together with the fitted curves. 
From these fits we can extrapolate the critical depth $p^\ast$.
The resulting values as a function of the problem size $n$ are reported in Figure~\ref{fig:pvsn}(b), where the dashed gray line indicates the region expected under the hypothesis of exponential growth ($\pm 1\sigma$). The critical depth is computed as the smallest integer $p$ satisfying
\[
p^\ast = \min \{ p \in \mathbb{Z}^+ : AR_{\text{exp}}(p) \ge AR^{\text{target}}_{\text{exp}} \}.
\]
Comparing the observed $p^\ast(n)$ with the null hypothesis of exponential scaling yields p-values below $0.05$, allowing us to reject exponential growth and supporting sub-exponential scaling. 
These findings align with previous empirical studies \citep{akshay2022circuit, wauters2020polynomial}, which reported approximately linear scaling of the QAOA depth with problem size to maintain a constant approximation ratio.

\vspace*{0cm} 

\begin{strip}
    \centering
    \includegraphics[width=0.9\linewidth]{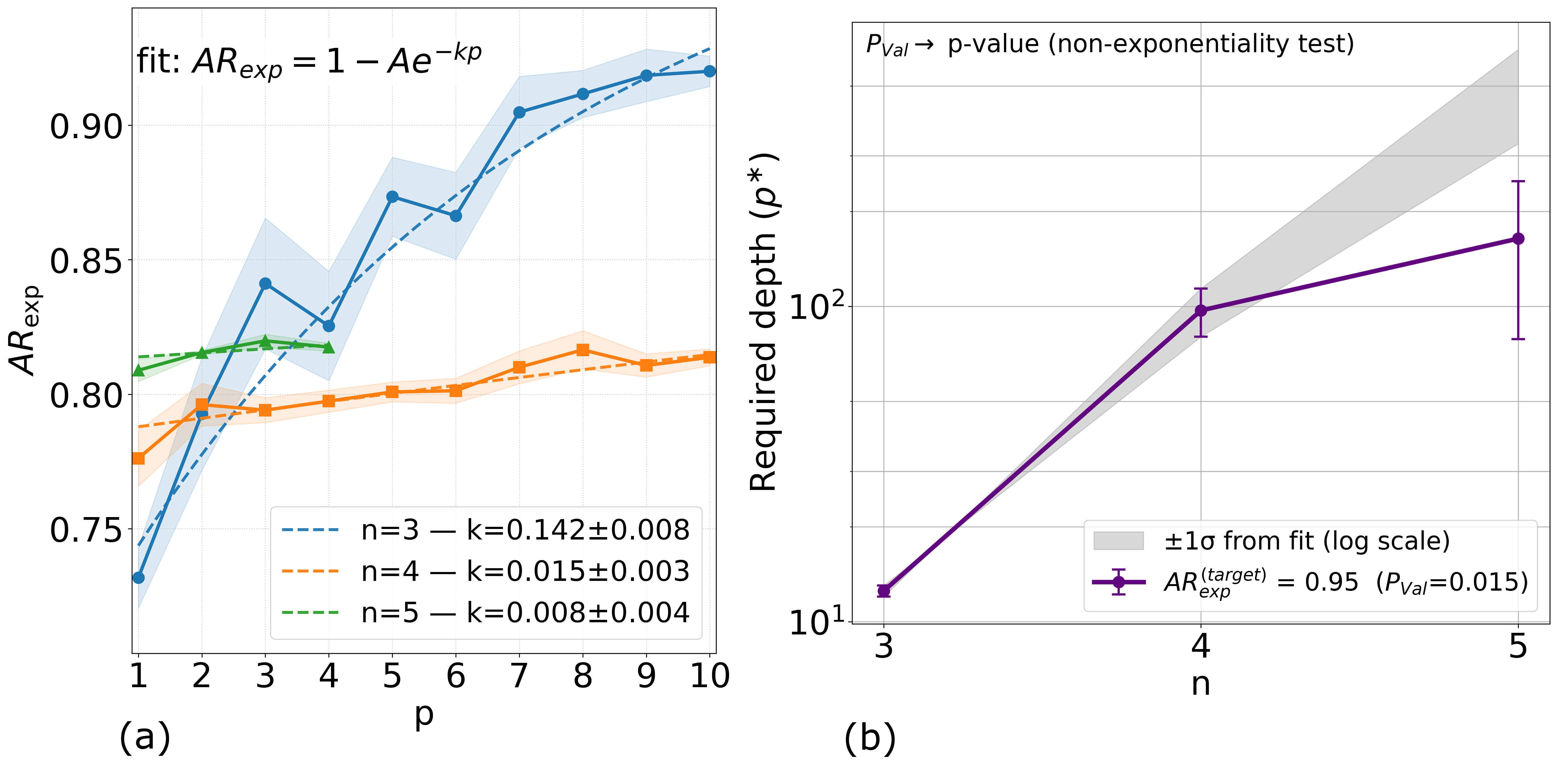}
    \captionof{figure}{Relationship between circuit depth and QAOA performance. 
    (a) Approximation ratio $AR_{\text{exp}}$ as a function of the circuit depth $p$, fitted using the exponential saturation model 
    $AR_{\text{exp}}(p) = 1 - A e^{-k p}$. The uncertainty on each data point is given by the standard error of the median. 
    (b) Estimated critical depth $p^\ast$ required to achieve a target approximation ratio, as a function of problem size $n$. 
    In panel (b), a $p$-value test was performed under the null hypothesis of exponential scaling of the data.}
    \label{fig:pvsn} 
\end{strip}
\FloatBarrier

\subsection{Analysis on Large Datasets: Cl-QAOA}
\label{subsec:large_datasets}

\begin{figure*}[!ht]
    \centering
    \includegraphics[width=0.9\linewidth]{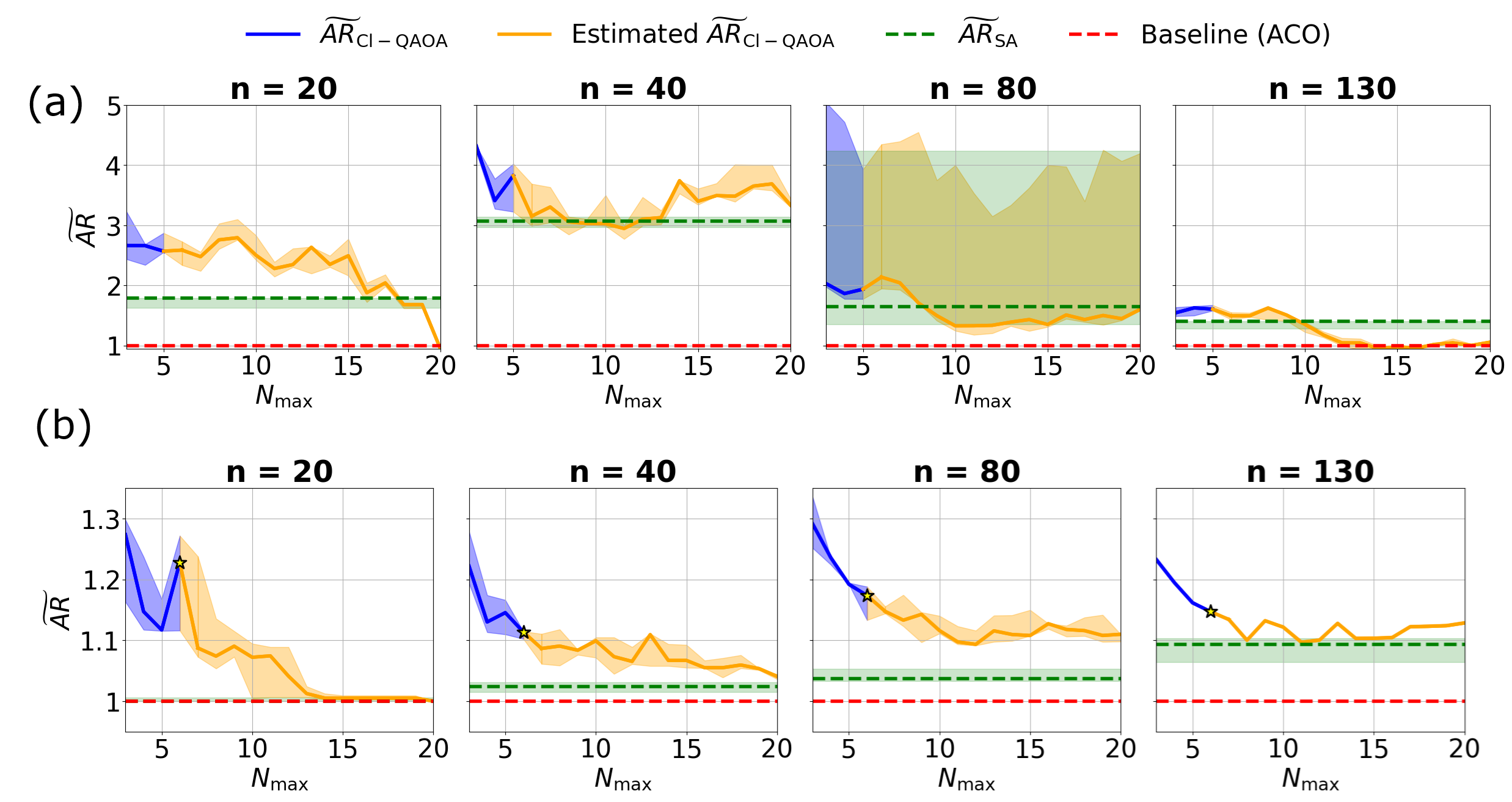}
    \caption{Comparison between synthetic (a) and realistic (b) datasets using the clustering QAOA approach (Cl-QAOA). The starred points in panel (b) correspond to $N_{\max}=6$ and were obtained using HPC resources at Cineca to handle large-scale computations.}
    \label{fig:AR_vs_cluster_size}
\end{figure*}

This section presents the analysis of large-scale TSP instances ($n>6$), which cannot be addressed directly using standard QAOA and instead require the Cl-QAOA algorithm. Figures~\ref{fig:AR_vs_cluster_size}(a) and (b) illustrate the effect of the maximum sub-problem size $N_{\text{max}}$ on the overall solution quality for the synthetic and the OSM datasets, respectively.

For the synthetic datasets, Cl-QAOA achieves solution quality comparable to SA for $n=40$, $80$, $130$. For $n=20$, Cl-QAOA matches SA performance when $N_{\text{max}}\ge 15$ and eventually surpasses it. Using ACO as a reference baseline, the same solution quality is reached for $n=20$ with $N_{\text{max}}=20$ and for $n=130$ with $N_{\text{max}}\ge 12$.

For the OSM Milan subway dataset, Cl-QAOA shows lower average performance. It matches the solution quality of SA for $n=20$ with $N_{\text{max}}\ge 13$ and for $n=130$ with $N_{\text{max}}\ge 8$. In comparison with ACO, the same solution quality is achieved only for $n=20$ with $N_{\text{max}}\ge 14$, indicating that matching ACO performance is more challenging within the tested range.

\newpage
\subsection{Time Complexity of Cl-QAOA}

\begin{figure}[!ht]
    \centering
    \includegraphics[width=0.76\columnwidth]{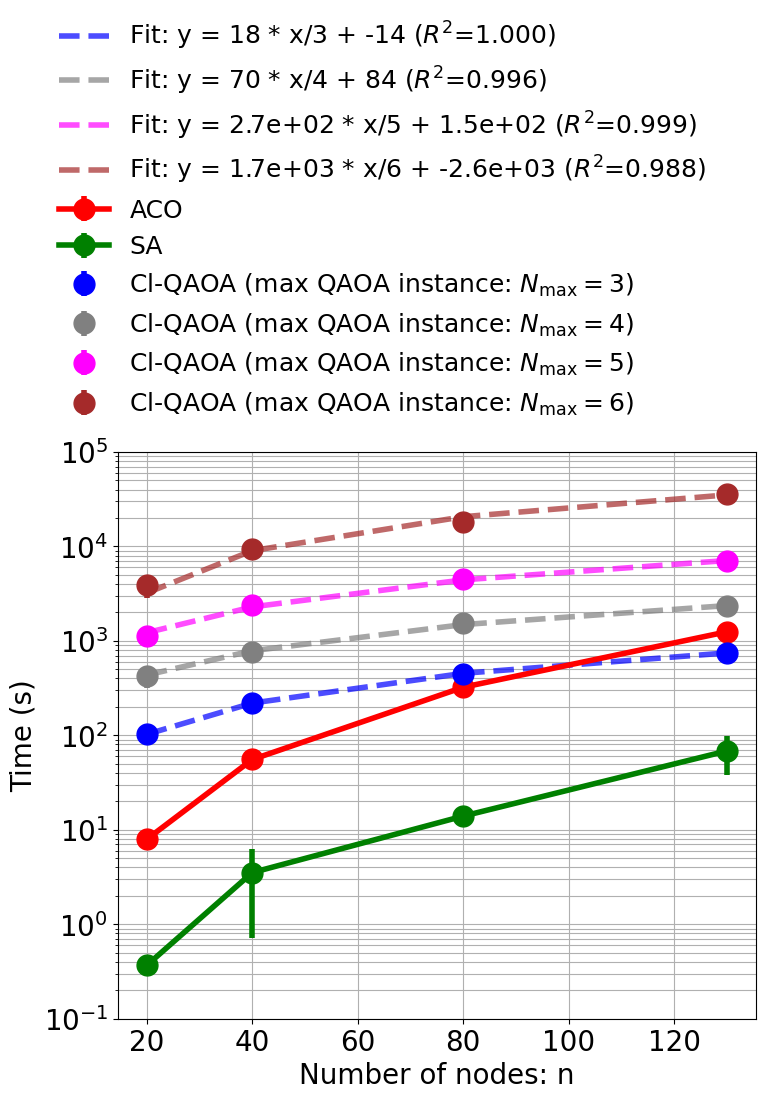}
    \caption{Execution time of Cl-QAOA compared to classical metaheuristics ACO and SA on the OSM Milano subway datasets. The x-axis indicates the problem size $n$, and the y-axis reports the runtime in seconds.}
    \label{fig:synthetic-vs-realistic_qaoa+clustering}
\end{figure}

The runtime of Cl-QAOA exhibits a linear dependence on the problem size $n$, as illustrated in Figure~\ref{fig:synthetic-vs-realistic_qaoa+clustering}. For each tested value of the maximum sub-problem size $N_{\max}$, fixing $N_{\max}$ results in a clear linear scaling with respect to $n$, supporting the model reported in Subsection~\ref{subsec:time_model}.

The figure also reports the computational time of the classical metaheuristics ACO and SA, both of which display polynomial growth. Notably, for $n \ge 100$, the execution time of ACO becomes higher than that of Cl-QAOA with $N_{\max}=3$. This crossover point highlights the scalability advantage of Cl-QAOA for larger instances, where classical methods begin to incur higher computational costs.

These results indicate that, beyond this threshold, the clustering QAOA approach offers a runtime advantage over ACO, making it a promising candidate for tackling large-scale combinatorial optimization problems in practical scenarios.

% \newpage
\section{Discussion}
\label{sec:discussion}

The analysis of the results is organized into two parts: the performance of QAOA on small TSP instances and the behavior of Cl-QAOA on larger problem sizes.

\subsection{QAOA}

The shot analysis for small TSP instances ($n \leq 6$) indicates that QAOA maintains stable performance across different problem sizes and constraint configurations. Notably, a circuit depth of $p = 1$ was sufficient to obtain optimal solutions in all tested cases, as reported in Tables~\ref{tab:shot_analysis_combined} and \ref{tab:shot_n6}.

The dependence on circuit depth further confirms that high-quality solutions can be achieved with small $p$, as illustrated in Fig.~\ref{fig:all_n4_n5_n6_AR}.
Results indicate that increasing $p$ generally improves $AR_{\text{exp}}$, particularly for smaller instances ($n = 3$ and $n = 4$), where modest depth increases yield noticeable gains. For $n = 5$ and $n = 6$, improvements are less pronounced, likely due to the exponential growth of the solution space. Nonetheless, the trend confirms that deeper circuits enhance solution quality, even when optimal solutions are already accessible at shallow depths, as shown in the shot analysis.

Since deeper circuits are more sensitive to noise and decoherence on current quantum devices, the ability to reach optimal or near-optimal performance with minimal depth is a substantial practical advantage. These findings support the suitability of QAOA for near-term quantum hardware, where circuit depth is a key limiting factor.

As shown in Fig.~\ref{fig:time_QAOA_vs_p} and consistent with previous studies \citep{weidenfeller2022scaling}, the total simulation time grows approximately quadratically with the circuit depth $p$. This trend reflects the increasing complexity of the optimization process: the number of classical iterations $I$ required for convergence scales linearly with $p$, as deeper circuits introduce more parameters and enlarge the search space, demanding additional iterations to reach convergence.

In addition, the scaling of the critical depth $p^\ast$, required to maintain a constant approximation ratio $AR_{\mathrm{exp}}$ with increasing problem size, exhibits sub-exponential scaling with $n$, as shown in Fig.~\ref{fig:pvsn}. Although no universal asymptotic law for the ideal depth $p$ is known, several studies suggest that at least a logarithmic scaling is necessary to surpass classical heuristics~\citep{farhi2020quantum,hastings2019classical,bravyi2020obstacles}.

According to the runtime model in Equation~\ref{eq:time_QAOA} and our analyses, QAOA for TSP does not appear to include exponential terms under the stated assumptions. This suggests that QAOA exhibits a scaling behavior that remains manageable when used as a probabilistic algorithm aimed primarily at lowering the expected value of the cost function.
If we consider a more deterministic scenario—requiring the sampling of solutions within a fixed cost threshold—the number of shots $s^\ast$ from Equation~\ref{eq:time_QAOA} could, in principle, grow rapidly. However, this growth strongly depends on how close the target cost is to the optimum and on the quality of circuit optimization. In practice, well-optimized circuits concentrate probability on low-cost regions, which can reduce $s^\ast$ from exponential to sub-exponential or even polynomial scaling. This means that, with effective parameter optimization and moderate thresholds, the sampling stage should remain tractable.

These considerations suggest that QAOA is not inherently exponential and that, with a well-tuned variational quantum circuit, even stricter sampling requirements can remain manageable. This points to a favorable outlook for applying QAOA to larger combinatorial problems as quantum hardware continues to advance.

%%%%%%%%%%%%%%%%%%%%%%%%%%%%%%%%%%%%%%%%%
\subsection{Cl-QAOA}

The clustering QAOA approach proposed in our work has shown promising results. As illustrated in Fig.~\ref{fig:AR_vs_cluster_size}, the relationship between $\widetilde{AR}_{\text{Cl-QAOA}}$ and the maximum sub-problem size $N_{\text{max}}$ is non-monotonic and exhibits noticeable fluctuations even for nearby values of $N_{\text{max}}$. This behavior is probably due to the clustering process: increasing the number of clusters does not always improve the solution, as splitting nodes that should remain connected can degrade performance. Nevertheless, the overall trend remains favorable: a larger $N_{\text{max}}$ generally leads to better solutions, although the improvement diminishes beyond a certain threshold. This conclusion is based on tests with $N_{\text{max}}$ up to 20: while solution quality in this range does not always match the ACO baseline, it is reasonable to expect that for larger values of $N_{\text{max}}$, Cl-QAOA could reach and potentially surpass ACO performance.

Moreover, the solution quality differs significantly between synthetic and OSM datasets when compared to the ACO baseline, with synthetic instances generally performing worse. This is likely because realistic datasets exhibit spatial and structural patterns that enable more effective clustering and coherent sub-problems, whereas synthetic datasets lack such structure, reducing clustering efficiency and overall solution quality.

As shown in Fig.~\ref{fig:synthetic-vs-realistic_qaoa+clustering}, the observed runtime confirms the linear scaling trend predicted by our model in Subsection~\ref{subsec:time_model}. This behavior suggests that, for large problem sizes, Cl-QAOA can achieve a computational advantage over ACO and SA, which typically scale as $\mathcal{O}(n^3)$ for real-world TSP instances~\citep{dorigo2004ant,kirkpatrick1983optimization}. In fact, even with $N_{\text{max}} = 3$, the advantage becomes evident starting from $n \approx 100$.
It is worth noting that on actual quantum hardware, the time required for executing the quantum component of the algorithm ($T_{\text{QAOA}}(N_{\text{max}})$ in Equation~\ref{eq:time_clqaoa}) would be drastically lower than the runtimes observed in the simulations conducted in this study. This reduction would not only amplify the temporal advantage for larger $N_{\text{max}}$ values but also enable the use of deeper circuits, improving solution quality while maintaining practical execution times.

In summary, Cl-QAOA serves as a practical bridge for applying QAOA to real-world problems even on current quantum hardware. Its structure naturally adapts to hardware improvements: as the availability of qubits increases, the algorithm can progressively reduce clustering, moving toward full QAOA by increasing $N_{\text{max}}$ up to the full problem size $n$. This flexibility makes Cl-QAOA an effective transitional strategy, enabling early adoption of quantum optimization while maintaining scalability for future devices.

\section{Conclusion and Future Work}
\label{sec:conclusions}

This work explored the application of the Quantum Approximate Optimization Algorithm (QAOA) to the Traveling Salesman Problem (TSP) under realistic constraints, including Binary Node Compatibility (BNC), road accessibility, and time-step limitations. We proposed a QUBO-based formulation capable of encoding these constraints without increasing quantum resource requirements, and introduced a Grover-inspired mixer to improve convergence toward feasible solutions.

Experiments on both synthetic and real-world datasets demonstrated that QAOA can consistently identify optimal solutions for instances up to $n = 5$ on a local workstation and up to $n = 6$ on the Cineca Leonardo HPC infrastructure. Remarkably, these results were achieved using a single QAOA layer and a limited number of measurement shots, regardless of constraint type, confirming that optimal performance can be obtained with shallow circuits. Our analysis of the approximation ratio $AR_{\text{exp}}$ revealed that increasing circuit depth $p$ leads to diminishing improvements as problem size grows, with constrained instances performing worse due to penalty-based constraint enforcement. 

A detailed runtime analysis shows that QAOA exhibits, at worst, sub-polynomial scaling with problem size during the optimization phase. Based on our proposed runtime scaling model, even the final sampling stage could remain sub-exponential once the variational parameters are well-optimized, as the circuit effectively concentrates probability on low-cost solutions.

To address larger instances, we introduced the clustering QAOA (Cl-QAOA) framework, which integrates classical clustering with QAOA. Tested on instances up to $n = 130$, Cl-QAOA achieved approximation ratios comparable to classical metaheuristics such as Ant Colony Optimization (ACO) and Simulated Annealing (SA), while maintaining linear scalability. Notably, performance improved more consistently on real-world datasets than on synthetic ones, indicating that structural properties that may be present in real-world networks can be exploited by Cl-QAOA to enhance solution quality. As quantum devices evolve, the progressive relaxation of clustering—enabled by increasing qubit counts—may allow Cl-QAOA to smoothly transition toward full QAOA, supporting the deployment of quantum optimization on real-world problem scales.

Future research will focus on extending the framework to support more general constraints, adapting it to complex routing problems such as the Vehicle Routing Problem (VRP), and exploring alternative encodings to reduce qubit requirements \citep{bako2025prog}. Enhancing the clustering phase using more ad-hoc methods tailored to clusters on graphs \citep{tsitsulin2023graph, wang2023overview} in Cl-QAOA may further improve decomposition quality and overall performance.
Moreover, testing the algorithm on real quantum hardware will provide valuable insights into noise effects, gate fidelities, and execution times, informing practical implementation strategies.

These directions aim to bridge the gap between theoretical quantum algorithms and their practical deployment in large-scale optimization, paving the way for quantum-enhanced solutions in real-world logistics and routing applications.
\appendix
\section{Classical Baselines: SA and ACO}
\label{appendix:classical}
\paragraph{Simulated Annealing (SA)}
We adopt a standard SA framework with logarithmic cooling and reheating. The initial temperature $T_0$ is set to the standard deviation of the cost matrix. At iteration $k$, a neighbor tour $S'$ is generated via swap, insert, or segment reversal moves and accepted with probability:
\[
P_{\rm accept} =
\begin{cases}
1, & \Delta \le 0,\\
\exp(-\Delta/T_k), & \Delta > 0,
\end{cases}
\]
where $\Delta$ is the cost difference and $T_k = T_0/(1+\ln(1+k))$. If no improvement occurs for $r=1\%$ of iterations, the temperature is reheated by multiplying by $1.2$. The total iterations scale as $K=100n^2$, and since each evaluation costs $O(n)$, the overall complexity is $O(n^3)$. Finally, a deterministic 2-opt local search refines the best tour.

\paragraph{Ant Colony Optimization (ACO)}
ACO follows the classical formulation with capped parameters for scalability. Each ant constructs a tour using transition probabilities:
\[
P_{i\to j} = \frac{\tau_{i,j}^\alpha (1/d_{i,j})^\beta}{\sum_{\ell\in\mathcal{U}} \tau_{i,\ell}^\alpha (1/d_{i,\ell})^\beta},
\]
with $\alpha=1.0$, $\beta=2.0+\log n$. Pheromone evaporation and reinforcement follow standard rules with bounds $[\tau_{\min},\tau_{\max}]=[0.01,5.0]$. We use $m=\min(2n,50)$ ants and $I=100+3n$ iterations, giving effective complexity $O(n^3)$ due to capped $m$. A final 2-opt refinement is applied as in SA.
\section{Scaling of $\langle H_C \rangle_{\gamma,\beta}$ and $\mathrm{Var}(H_C)_{\gamma,\beta}$}
\label{appendix:scaling}

The QUBO cost Hamiltonian for TSP with logistical constraints is:
\[
H_C = \sum_{i,j=1}^{n} \sum_{t=1}^{n-1} \tilde{\omega}_{i,j} x_{i,t} x_{j,t+1} 
\]
\[
+ \lambda_P \sum_i \Big(\sum_t x_{i,t} - 1\Big)^2 
+ \lambda_T \sum_{i,t} T_{i,t} x_{i,t},
\]
with $x_{i,t} \in \{0,1\}$ and and $\tilde{\omega}_{i,j}$ the cost of moving from node $i$ to node $j$ adding, if necessary, logistical constraints (see Equation \ref{eq:omegatilde}). Binary variables are mapped to Pauli operators via:
\[
x_{i,t} \mapsto \frac{1 - Z_{i,t}}{2}, \qquad
\]
\[
x_{i,t} x_{j,t+1} \mapsto \frac{1}{4}(1 - Z_{i,t} - Z_{j,t+1} + Z_{i,t} Z_{j,t+1}).
\]
Each quadratic term generates up to 4 Pauli terms, so the Hamiltonian expands to:
\[
H_C = \sum_{k=1}^{\eta} c_k \sigma_k, \qquad \eta = \mathcal{O}(n^4).
\]
Assuming independent measurements:
\[
\langle H_C \rangle_{\gamma,\beta} = \sum_{k=1}^{\eta} c_k \langle \sigma_k \rangle = \mathcal{O}(n^4),
\]
\[
\mathrm{Var}(H_C) = \sum_{k=1}^{\eta} c_k^2 \mathrm{Var}(\sigma_k) \le \mathcal{O}(n^8),
\]
since $\mathrm{Var}(\sigma_k) \le 1$.

\section*{Acknowledgements}
The authors acknowledge CINECA for providing the computational resources used to perform the simulations for the $n = 6$ instances. This work was supported by the Italian National Centre for HPC, Big Data and Quantum Computing, Spoke 10 “Quantum Computing” (\url{https://www.supercomputing-icsc.it/spoke-10-quantum-computing/}).

%% The declaration on generative AI comes in effect
%% in Janary 2025. See also
%% https://ceur-ws.org/GenAI/Policy.html
%\section*{Declaration on Generative AI}
 %During the preparation of this work, the author(s) used ChatGPT-4 in order to: Grammar and spelling check. After using these tool(s)/service(s), the author(s) reviewed and edited the content as needed and take(s) full responsibility for the publication's content. 

%%===========================================================================================%%
%% If you are submitting to one of the Nature Portfolio journals, using the eJP submission   %%
%% system, please include the references within the manuscript file itself. You may do this  %%
%% by copying the reference list from your .bbl file, paste it into the main manuscript .tex %%
%% file, and delete the associated \verb+\bibliography+ commands.                            %%
%%===========================================================================================%%

\bibliographystyle{unsrtnat}
\bibliography{sn-bibliography}% common bib file

%% if required, the content of .bbl file can be included here once bbl is generated
%%\input sn-article.bbl

\end{document}